\documentclass[a4paper, onecolumn]{article}
\usepackage[dvips]{graphicx}
\begin{document}
\title{Advanced Satellite-based Frequency Transfer\\ at the 10\(\bf ^{-16}\) Level}
\author{M. Fujieda, S-H. Yang, T. Gotoh, S-W. Hwang, H. Hachisu, H. Kim, Y. K. Lee,\\ R. Tabuchi, T. Ido, 
W-K. Lee, M-S. Heo, C. Y. Park, D-H. Yu, and G. Petit
\thanks{M. Fujieda, T. Gotoh, H. Hachisu, R. Tabuchi, T. Ido are with the
National Institute of Information and Communications Technology, Koganei, Tokyo, 
184-8795 Japan. e-mail: miho@nict.go.jp.}
\thanks{S-H. Yang, S-W. Hwang, Y. K. Lee, H. Kim, W. K. Lee, M-S. Heo, C. Y. Park, D-H. Yu 
are with the Korea Research Institute of Standards and Science, Daejeon, 305-600 Korea.}
\thanks{G. Petit is with the Bureau International des Poids et Mesures, S\`evres F-92312 France.}}
\maketitle
\begin{abstract}
Advanced satellite-based frequency transfers by TWCP and IPPP have been 
performed between NICT and KRISS. 
We confirm that the disagreement between them is less than 
\(\bf 1\times10^{-16}\) at an averaging time of several days. 
Additionally, an intercontinental frequency ratio measurement of 
Sr and Yb optical lattice clocks was directly performed by TWCP. 
We achieved an uncertainty at the mid-10\(\bf ^{-16}\) 
level after a total measurement time of 12 hours. 
The frequency ratio was consistent with the recently reported values within the uncertainty. 
\end{abstract}
%
%
\section{Introduction}
%
%
Satellite-based 
time and frequency transfers are in demand for intercontinental links and typically utilize 
pseudorange measurements using a code phase of a signal modulated by a pseudorandom 
noise (PN) sequence. Increasing the chip rate of the PN sequence improves the measurement 
precision, although it occupies a wider frequency bandwidth at the same time. 
This has limited the measurement precision of two-way satellite time and frequency transfer 
(TWSTFT) to an insufficient level for the comparison of advanced optical clocks. 
To improve the precision, the National Institute of Information and Communications Technology (NICT) 
has developed a two-way carrier-phase satellite frequency transfer (TWCP) [1]. 
It achieves sub-picosecond-level precision, which is three orders of magnitude better than that of TWSTFT [2]. 
On the other hand, the precise point positioning (PPP) technique utilizing code and carrier phases has been 
used by GPS time and frequency transfer (hereafter called GPS transfer) and contributed to the 
International Atomic Time (TAI) computation [3, 4]. 
The measurement precision is improved by two orders of magnitude to a few tens of ps 
by the application of the carrier phase in the GPS transfer. 
However, solving the phase ambiguity may introduce random errors in GPS transfer results. 
To overcome this limitation, the integer PPP (IPPP) technique has been developed [5]. 
It has been applied to GPS transfer and recently demonstrated a \(1\times10^{-16}\) 
frequency transfer accuracy [6]. Thus, these advanced satellite-based frequency transfer techniques 
such as TWCP and IPPP have the potential to enable optical clock comparisons in a very long baseline. 
Aiming at the evaluation and comparison of their techniques, NICT and the Korea Research Institute 
of Standards and Science (KRISS) established the TWCP link in December 2016. 
In this paper, we describe the frequency transfer techniques in Sec. II and introduce 
the comparison between the two techniques in Sec. III. In Sec. IV, the frequency ratio 
measurement of Sr and Yb optical lattice clocks by TWCP is demonstrated. 
\section{Frequency Transfer Techniques and Setups}
NICT and KRISS have performed TWCP and GPS transfer on a regular basis. 
Here, we introduce the TWCP and IPPP techniques briefly. 
Then their measurement setups at NICT and KRISS are shown.
\subsection{TWCP}
By two-way signal exchange, the delay terms in the propagation path are 
almost canceled in TWCP. When we determine a pseudorange from the 
carrier phase of the transmitted signal from an earth station A at an earth station B, 
however, we have to remove two terms: the Doppler effects caused by the satellite 
motion and the phase noise induced by an onboard oscillator in frequency 
conversion from uplink to downlink frequencies because most of communication 
satellites do not carry an atomic clock. It was shown in [2] that the 
mathematical cancellation by using four carrier phases of the four signals 
from A to A, from A to B, from B to A, and from B to B is effective in removing them. 
Additionally, the ionosphere delay is not canceled either because 
the uplink and downlink frequencies are different. 
Therefore, we utilize an ionosphere map and compute the delays using 
the total electron contents (TECs) over the earth stations. 
Since the carrier phase is continuously tracked and accumulated in TWCP 
measurement without any fixation of phase ambiguity, 
the result basically keeps the continuity as long as the measurement continues. \\
~~Figures 1 (a) and (b) show schematics of the earth stations at NICT and KRISS, 
respectively. We use an arbitrary waveform generator for the signal generation 
and an analogue-to-digital (A/D) sampler for carrier-phase detection. 
The TWCP signal has a frequency bandwidth of 200 kHz. 
The code phase of 128 k chips per second (cps) is detected, as well as the carrier phase, 
and it helps the signal tracking. The frequency conversion is carried out 
by commercial frequency up- and downconverters from 70 MHz 
to uplink and downlink frequencies of 14 and 11 GHz, respectively. 
The reference signals of 10 MHz are provided to the frequency converters 
to maintain the carrier phase coherence. At NICT, a dedicated earth station 
is used for the TWCP measurement. 
On the other hand, at KRISS, the TWCP and conventional TWSTFT 
measurements share one earth station. 
The two signals for the TWCP and TWSTFT are combined and divided 
at 70 MHz by a signal combiner and divider, respectively. 
Figure 1 (c) shows the reception spectrum where the same center frequency 
was used by the TWCP signal and the TWSTFT signal with a bandwidth of 2.5 MHz. 
We did not observe any interference between them caused by sharing 
the same frequency bandwidth. 
We started continuous TWCP measurement alongside TWSTFT in December 
2016 using a geostationary satellite named Eutelsat 172A. 
For the ionosphere delay correction between NICT and KRISS, 
we use a global ionosphere map produced 
by the Center for Orbit Determination in Europe (CODE) [7]. 
The typical amplitude of the ionosphere delay is about 10 ps.
%
\begin{figure}[p]
\begin{center}
\vspace*{-5mm}
\hspace*{-5mm}
\includegraphics[width=100mm]{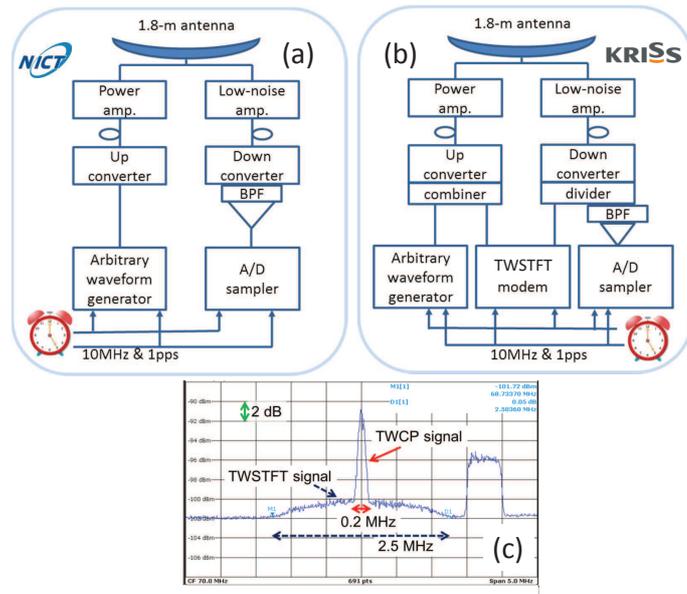}
\vspace*{-50mm}
\caption{Schematics of the earth stations at (a) NICT and (b) KRISS. 
BPF: band-pass filter, amp.: amplifier, A/D: analogue-to-digital converter. 
(c)  Spectrum of the reception signals for TWCP and TWSTFT.}
\end{center}
\end{figure}
\subsection{IPPP}
In IPPP, as in classical PPP, the user's clock is determined 
from dual frequency GPS phase measurements using 
satellite clock products generated by analysis of a global network. 
For IPPP, the satellite products generated by the GRG analysis centre [8] 
are designed so that the user can determine the phase ambiguities 
for each satellite pass and the two frequencies as integers N1/N2. 
This is done in a two-step procedure with the CNES GINS software, 
first determining the widelane ambiguity Nw = N2 - N1 and then 
determining e.g. N1, see details in [6, 8]. 
Clock differences are then continuous as long as there is no discontinuity 
in the set of integer ambiguities for all satellite passes. \\
~~As shown in [6], the above treatment is performed on a station by station basis, 
independently for each day, and the continuity between successive daily batches 
is then to be established. By design, discontinuities between batches 
should be an integer number of the so-called narrowlane 
wavelength \(\lambda_{N}\) (\(\sim\)350 ps) and it is simpler 
to determine these discontinuities when forming 
the difference of two station clocks, i.e. a time link. Indeed, 
when the instability of the two compared clocks is sufficiently low, 
the extrapolation noise from batch to batch is much lower than \(\lambda_{N}\) 
and it is easy to determine the discontinuities as an integer number of \(\lambda_{N}\). 
This extrapolation technique was used for the IPPP solution 
between UTC(NICT) and UTC(KRIS) which are both based on H-masers. 
Note that such discontinuities must also be determined when all satellite 
ambiguities are reset within a daily batch, e.g. by a short data gap.\\
~~The GPS receivers used at NICT and KRISS are Septentrio PolaRx 4 
and Ashtech Z12T connected to the reference signals from UTC(NICT) and UTC(KRIS), respectively.
\section{Comparison between TWCP and IPPP}
%
\begin{figure}[p]
\begin{center}
\includegraphics[width=85mm]{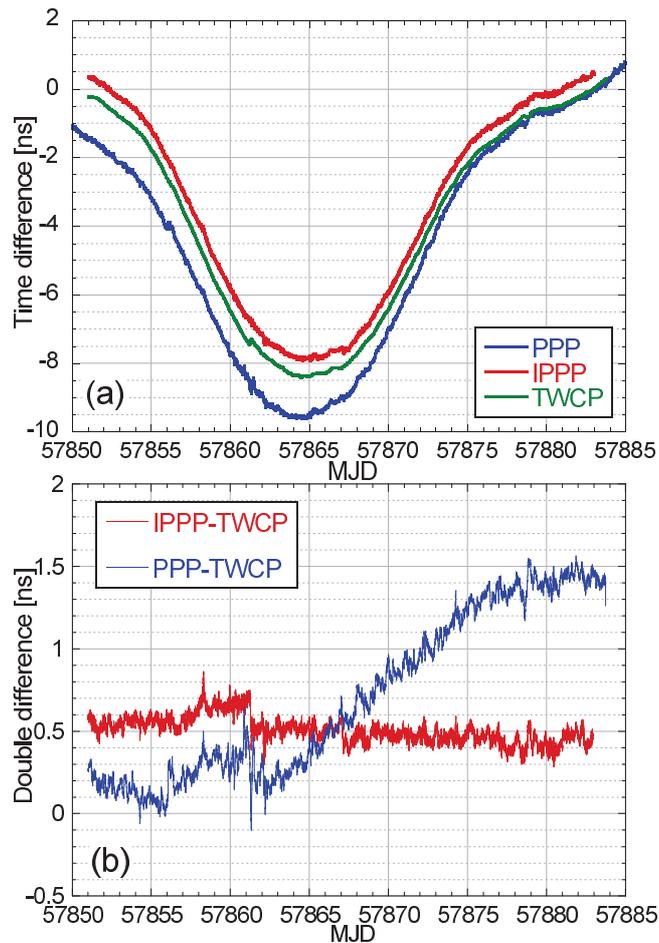}
\caption{ (a) Time difference between UTC(NICT) and UTC(KRIS), 
(b) double differences of IPPP-TWCP and PPP-TWCP.}
\end{center}
\end{figure}
We compared the measurement results obtained by PPP, IPPP, and TWCP 
for two periods: (1) from MJD 57772 to MJD 57784 and (2) from MJD 57851 to MJD 57883. 
Here, PPP is computed with NRCan PPP software using 15 days batch and IGS final products. 
The TWCP measurements were continuously carried out without any downtime. 
The measurement rates of PPP, IPPP, and TWCP are 300, 30, and 1 s, respectively. 
Figure 2 (a) shows the time difference between UTC(NICT) and UTC(KRIS) for period (2). 
For better visibility, offset values are inserted. 
Figure 2 (b) shows the double differences of IPPP-TWCP and PPP-TWCP. 
A time jump can be observed at MJD 57861. It is clear that the TWCP result 
causes it because the signal-to-noise ratio at the KRISS station suddenly 
decreased by 8 dB owing to heavy rain, and a phase excursion of 0.2 ns 
occurred in the TWCP result. Additionally, a small jump can be seen at MJD 57856. 
It was found that the GPS receiver at NICT caused a reset of 
all ambiguities at that time. While IPPP found an exact integer number of cycles 
to go through the reset, PPP could not. Furthermore, the double difference 
of PPP-TWCP indicates a clear gradient, which can also be seen in the result for period (1). 
We assume that the gradients show disagreements among the techniques, 
and they are summarised in Table 1. IPPP and TWCP show consistency at the \(10^{-17}\) level.
Figure 3 shows the modified Allan deviation of UTC(NICT)-UTC(KRIS) for period (2). 
The stabilities longer than 1 day are limited by those of the time scales. 
The double differences of IPPP-TWCP and PPP-TWCP are free from the limitation. 
While the curve of IPPP-TWCP is decreasing and reaches \(10^{-17}\) level after 500,000 s, 
that of PPP-TWCP becomes flat. For period (1), similar stabilities are achieved of \(5.5\times10^{-17}\) 
at 250,000 s for IPPP-TWCP and \(1.7\times10^{-16}\) at 250,000 s for PPP-TWCP. 
By the comparisons among PPP, IPPP, and TWCP, it proved that TWCP, 
as well as IPPP, has superior long-term stability. 
Until now, unless the signal-to-noise ratio decreases suddenly 
owing to the weather conditions, the measurement has successfully continued and 
phase continuity can be preserved in the TWCP results.\\
~~Figure 4 shows the time differences of UTC(NICT)-UTC(KRIS) free from 
the time-scale variation, from which its two-day moving average is subtracted 
and then the remainder is averaged over 1 hour. 
As shown in Figure 4 (a), IPPP and PPP indicate some variations 
with periods of one day and half a day. Examination of each receiver's data 
shows that both receivers have one-day-period components. 
In Figure 4 (b), the time difference for TWCP is depicted together 
with the outdoor temperature at NICT. 
Since the diurnal change of outdoor temperature at KRISS is similar to that of NICT, 
it is not shown. The fluctuation in TWCP is smaller than those of IPPP and PPP. 
However, TWCP also shows one-day-period variation with an amplitude of 10 ps order. 
The ionosphere delays have already been compensated. 
The variation sometimes shows a weak inverse correlation with the outdoor temperature, 
for example, around April 25, 2017. 
Additionally, the variation around April 20, 2017 seems to have a positive correlation. 
We recently moved the outdoor power amplifier and low-noise amplifier 
into a temperature-stabilized box at the NICT earth station. 
Further analysis of the effect will be carried out in the near future. 
%
\begin{figure}[p]
\begin{center}
\hspace*{-2mm}
\includegraphics[width=90mm]{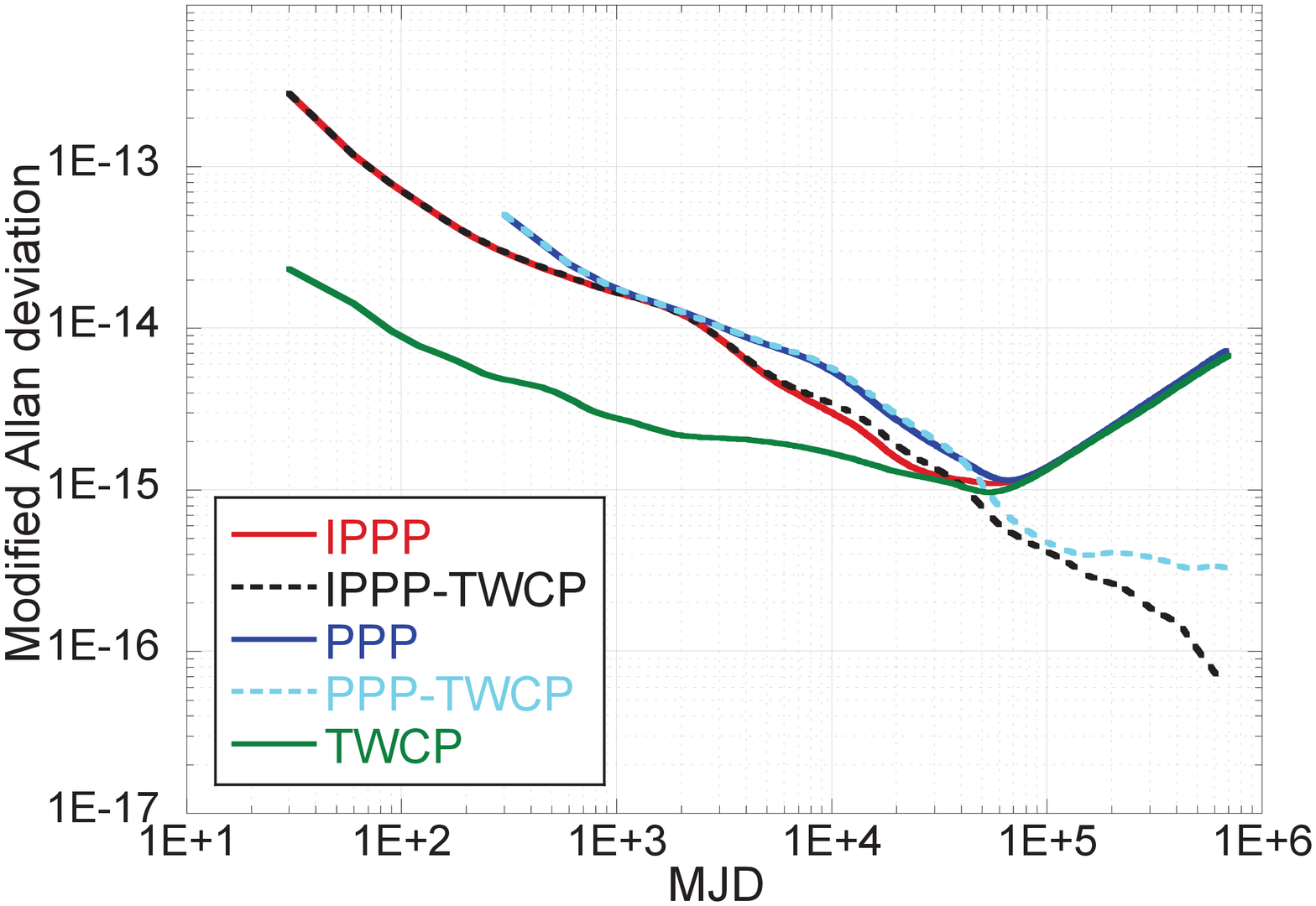}
\vspace*{-8mm}
\caption{Modified Allan deviation of UTC(NICT)-UTC(KRIS) from MJD 57851 to 57883.}

\end{center}
\end{figure}
%
\begin{table}[p]
\caption{Disagreement between PPP, IPPP and TWCP}
\begin{center}
\begin{tabular}{c|c|c}
\hline \hline
Period & Difference & Disagreement (\(10^{-16}\)) \\ \hline
(1) & IPPP-TWCP & -0.43 \\
MJD 57772-57784 & PPP-TWCP & 3.8 \\ \hline
(2) & IPPP-TWCP & -0.66 \\ 
MJD 57851-57883 & PPP-TWCP & 5.9 \\ \hline \hline
\end{tabular}
\end{center}
\end{table}
%
\begin{figure}[p]
\begin{center}
\includegraphics[width=85mm]{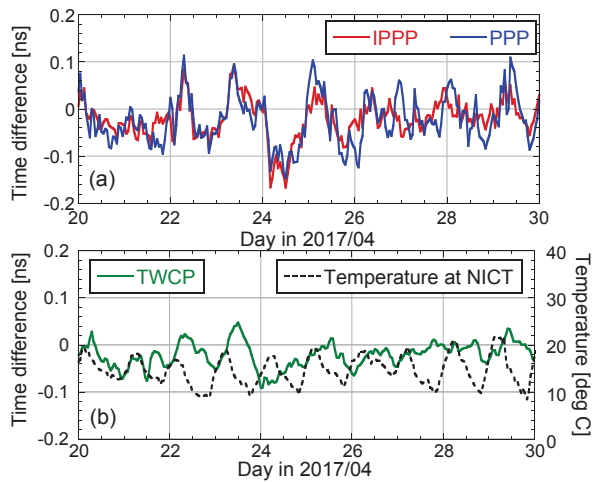}
\vspace*{-50mm}
\caption{Time difference free from time-scale variation: (a) IPPP and PPP, 
(b) TWCP and outdoor temperature at NICT.}
\end{center}
\end{figure}
\section{Yb/Sr Frequency ratio measurement by TWCP}
\subsection{Measurement Setup}
NICT and KRISS operate a \(^{87}\)Sr optical lattice clock [9, 10] and 
an \(^{171}\)Yb optical lattice clock [11]. 
The frequency ratio was measured by TWCP through microwave references. 
The optical clocks were continuously operated for around 4 hours per day over 3 days 
during February 1-3, 2017. We performed two local measurements and one TWCP 
measurement at the same time. The frequencies of optical clocks were measured with 
reference to the microwave references by using an optical frequency comb at each site. 
The systematic frequency shifts of the optical clock were corrected including the gravitational redshift. 
TWCP evaluates the frequency difference between the microwave references of both sites. 
At NICT and KRISS, a hydrogen maser, HM(NICT), and UTC(KRIS) 
with frequency \(\it f_{\rm HM(NICT)}\) and \(\it f_{\rm UTC(KRIS)}\), respectively, 
were used as the microwave references. 
These signals were provided to the earth stations for TWCP. 
By obtaining fractional frequencies \(\it y_{\rm Yb}\) and \(\it y_{\rm Sr}\) 
against each result of previous absolute frequency measurements 
\(\it\overline{f_{\rm Yb}}\) and \(\it\overline{f_{\rm Sr}}\) [11, 10] 
on the basis of \(\it f_{\rm UTC(KRIS)}\) and \(\it f_{\rm HM(NICT)}\), 
the frequency ratio of the two 
optical clock transitions \(\it \nu_{\rm Yb}/\it \nu_{\rm Sr}\) is measured as

\[
\frac{\it \nu_{\rm Yb}}{\it \nu_{\rm Sr}} =
\frac{\it y_{\rm Yb}}{\it y_{\rm Sr}} \cdot 
\frac{\it \overline{f_{\rm Yb}} }{\it \overline{f_{\rm Sr}} } \nonumber \]
\[
= \frac{\it f_{\rm HM(NICT)}}{\it f_{\rm Sr} / \overline{\it f_{\rm Sr}}  } \cdot 
\frac{\it f_{\rm UTC(KRIS)}}{\it f_{\rm HM(NICT)}} \cdot
\frac{\it f_{\rm Yb} / \overline{\it f_{\rm Yb}} }{\it f_{\rm UTC(KRIS)}} \cdot
\frac{\it \overline{f_{\rm Yb}} }{\it \overline{f_{\rm Sr}} }  (1)
\]

\subsection{Data analysis}
%
%
\begin{figure}[p]
\hspace*{-8mm}
\vspace*{-5mm}
\begin{center}
\includegraphics[width=80mm]{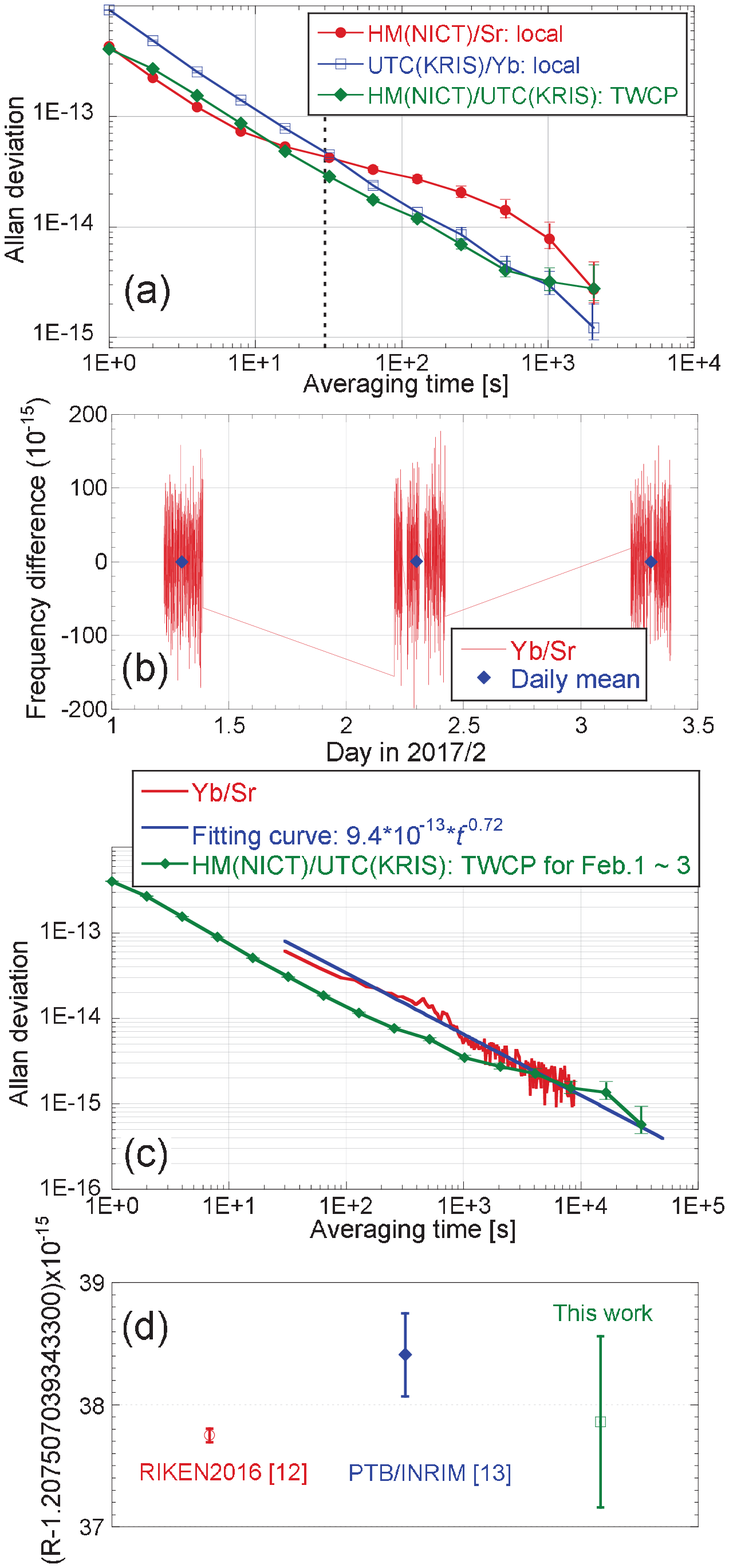}
\vspace*{-2mm}
\caption{(a) Allan deviations of two local measurements and one TWCP measurement 
carried out on Feb. 1. (b) Frequency difference of \(\it y_{\rm Yb}/\it y_{\rm Sr}-1\)
for 3 days. (c) Lumped Allan deviation of \(\it y_{\rm Yb}/\it y_{\rm Sr}\), 
its fitting curve and the Allan deviation of 
\(\it f_{\rm HM(NICT)}/\it f_{\rm UTC(KRIS)}\)  measured from Feb. 1-3. 
(d)  Frequency ratio R reported so far. }
\end{center}
\end{figure}
The data acquisition rates of the two local measurements and 
one TWCP measurement were 1 point per every second. 
First, we extracted the data where both optical clocks were simultaneously operated. 
Figure 5 (a) depicts their Allan deviations measured on Feb. 1. 
The HM(NICT) signal was transferred by an unfixed coaxial cable to the Sr clock at this measurement, 
which caused an unwanted fluctuation and instability at around 100 s. 
Since the stabilities shown in Figure 5 (a) display similar values around 30 s, 
we averaged three frequency ratios over 30 s, aligned the time stamps, and then calculated 
\(\it y_{\rm Yb}/\it y_{\rm Sr}\) following (1). 
The computed difference relative to 
\(\it\overline{f_{\rm Yb}}/\it\overline{f_{\rm Sr}}\) 
is depicted in Figure 5 (b). The Allan deviation was calculated from the lumped data 
for 3 days as shown in Figure 5 (c) by a red line, where the fitting curve of 
\(9.4\times10^{-13}\times t^{-0.72}\) is also depicted by a blue line. 
Table 2 summarises the daily mean values. Weighting the daily mean values 
by the number of 30-s points divided by the square of the daily standard deviation, 
we concluded the weighted frequency difference of 
\(4.9\times10^{-16}\) for the 3-day measurement. 
As for the daily statistical uncertainty, it was computed using the fitting curve 
of the Allan deviation for the measurement period.
Table 3 shows the uncertainty budget. 
The total statistical uncertainty was determined using the mean of 
the daily statistical uncertainties divided by the square root of three: 
\( ((9.7^{2}+9.1^{2}+9.6^{2})/3/3)^{0.5}\times10^{-16}\).
In Figure 5 (c), the Allan deviation of \(\it f_{\rm HM(NICT)}/\it f_{\rm UTC(KRIS)}\) 
measured by TWCP for 3 days is depicted by the green line, 
which meets the fitting curve at \(5\times10^{-16}\) around 40,000 s. 
This implies that the total statistical uncertainty was estimated appropriately. 
The systematic uncertainty for TWCP was \(1\times10^{-16}\) 
from the disagreement with IPPP. On the other hand, the Sr and Yb lattice clocks 
have systematic uncertainties of \(0.5\times10^{-16}\) and 
\(1.2\times10^{-16}\), respectively. 
The uncertainties of the differential gravitational redshift between both clocks is 
\(0.4\times10^{-16}\). 
We achieved a total uncertainty of  \(5.8\times10^{-16}\). 
Thus, the consistency of the previous absolute frequencies reported by NICT and KRISS [10, 11] 
was confirmed by \(\it y_{\rm Yb} / \it y_{\rm Sr} \rm -1=(4.9\pm5.8)\times10^{-16}\). 
We concluded the frequency ratio, R, as \\
~~~~\(\rm R = \it\nu_{\rm Yb} / \it\nu_{\rm Sr} \rm = 1.207,507,039,343,337,86 (70) \). \\
The recently reported values are shown in Figure 5 (d) [12, 13]. 
We confirm that our result is consistent with them within the uncertainty.
%
\begin{table}[p]
\caption{Daily Mean Values}
\begin{center}
\begin{tabular}{l|c|c|c}
\hline \hline
 & Feb. 1 & Feb. 2 & Feb. 3 \\ \hline
\(\it y_{\rm Yb} / \it y_{\rm Sr}-1\) (\(10^{-15}\)) & 0.33 & 1.00 & 0.17 \\
Std. Dev. (\(\sigma\))(\(10^{-15}\)) & 57.60 & 60.13 & 54.85 \\
No. of 30-s points (N) & 479 & 517 & 477 \\
Measurement period (s) & 14370 & 15510 & 14310 \\
Daily statistical uncertainty (\(10^{-15}\)) & 0.97 & 0.91 & 0.96 \\ \hline
Weighted mean (\(10^{-15}\)) & \multicolumn{3}{c}{0.49} \\ \hline \hline
\end{tabular}
\end{center}
\end{table}
%
\begin{table}[p]
\caption{Uncertainty Budget}
\begin{center}
\begin{tabular}{l|c}
\hline \hline
 & (\(10^{-16}\)) \\ \hline
Statistical & 5.5 \\
Sr systematic & 0.5 \\
Yb systematic & 1.2 \\
Gravitational redshifts & 0.4 \\
Link systematic & 1.0 \\ \hline
Total & 5.8 \\ \hline\hline
\end{tabular}
\end{center}
\end{table}
\section{Conclusion}
We evaluated measurement results achieved by the advanced satellite-based 
frequency techniques of PPP, IPPP, and TWCP in the NICT-KRISS link. 
While the disagreement of PPP-TWCP remains at the \(10^{-16}\) level, 
that of IPPP-TWCP reaches the \(10^{-17}\) level at over 
\(5\times10^{5}\) s average time.\\
~~The intercontinental frequency ratio measurement of Sr and Yb optical lattice 
clocks was directly performed by TWCP. We performed a dead-time-free and 
simultaneous measurement for about 12 hours and confirmed 
that the achieved frequency ratio is consistent with those shown in other reports 
within a total uncertainty at the mid-\(10^{-16}\) level. 
In conclusion, not only optical fiber links but also advanced satellite-based frequency 
links are applicable to optical clock comparisons. 
The satellite-based techniques have the potential to realize uncertainty 
at the \(10^{-17}\) level when optical clocks are continuously operated for a week or more.
\section*{Acknowledgement}

The authors would like to thank H. Takiguchi now in JAXA for his technical support in PPP. The use of the GINS software from the CNES and of the GPSPPP software from the NRCan is gratefully acknowledged.

%




%
\end{document}